\documentclass[aps,8pt,prd,twocolumn,preprintnumbers,amsmath,amssymb,nofootinbib,superscriptaddress,a4paper]{revtex4}
\pdfoutput=1
\usepackage[colorlinks=true, pdfstartview=FitV, linkcolor=blue, citecolor=red, urlcolor=magenta]{hyperref}
\usepackage[utf8]{inputenc}
\usepackage{graphicx}
\usepackage{latexsym}
\usepackage{amsmath}
\usepackage{amsfonts}
\usepackage{amssymb}
\usepackage{verbatim}
\usepackage{dcolumn}
\usepackage{amssymb}
\usepackage{bm}
\usepackage{float}
\usepackage{subfigure}
\usepackage[english]{babel}
\usepackage{hyperref}

\newcommand{\be}{\begin{equation}}
\newcommand{\ee}{\end{equation}}
\newcommand{\bea}{\begin{eqnarray}}
\newcommand{\eea}{\end{eqnarray}}

\begin{document}

\title{Insight into the Microstructure of FRW Universe from a $P$-$V$ Phase Transition}


\author{Haximjan Abdusattar}
\email{axim@nuaa.edu.cn}
\affiliation{School of Physics and Electrical Engineering, Kashi University, Kashi 844009, Xinjiang, China}
\affiliation{College of Physics, Nanjing University of Aeronautics and Astronautics, Nanjing, 211106, China}
\affiliation{Key Laboratory of Aerospace Information Materials and Physics (NUAA), MIIT, Nanjing 211106, China}

\begin{abstract}

The $P$-$V$ phase transition of the FRW (Friedmann-Robertson-Walker) universe with a perfect fluid has recently been investigated, revealing that the four critical exponents near the critical point are consistent with the values predicted by mean field theory. Notably, the coexistence phase of the $P$-$V$ phase transition in the FRW universe above the critical temperature, which distinguishes it from van der Waals system and most of AdS black holes system. This unique property allows us to investigate the microstructure of the FRW universe as a thermodynamic system. Our analysis of the Ruppeiner geometry for the FRW universe reveals that the behavior of the thermodynamic scalar curvature near criticality is characterized by a dimensionless constant identical to that of the van der Waals fluid. Additionally, we observe that while repulsive interactions dominate for the coexistence samll phase with higher temperature, the scalar curvature for the coexistence large phase is always negative, indicating attractive interactions, providing new insights into the nature of interactions among the perfect fluid matter constituents in the expanding FRW universe.

\end{abstract}


\maketitle


\section{Introduction}

Phase transitions are fascinating phenomena that play a pivotal role in black hole thermodynamics and have been widely studied in recent years. In exploring phase transitions and critical phenomena, the thermodynamic pressure plays an important role. For black holes in AdS space, the cosmological constant is usually treated as the thermodynamic variable analogous to the pressure $P:=-\Lambda/(8\pi)$ \cite{Kastor:2009wy,Dolan:2010ha}, and its conjugate quantity is the thermodynamic volume $V$, which yields the presence of a pressure-volume term in the first law of thermodynamics. With such an identification, Ref.\cite{Kubiznak:2012wp} constructed an equation of state $P=P(V, T)$ for four-dimensional charged AdS black holes and found similar $P$-$V$ critical behaviors to those of the van der Waals fluid in ordinary thermodynamics. As a result, their critical exponents are the same as well. Subsequent studies have explored the critical behavior of black holes in numerous models with AdS background \cite{Hu:2018qsy,Bhattacharya:2017hfj,Bhattacharya:2017nru,Hendi:2012um,Hendi:2017fxp} (see also \cite{Gunasekaran:2012dq,Wei:2012ui,Cai:2013qga,Dehghani:2014caa,Xu:2015rfa,Cai:2014znn,Spallucci:2013osa,Majhi:2016txt,
Cheng:2016bpx,Dehyadegari:2017hvd,Kumara:2019xgt,Li:2020xkh,Abdusattar:2023fdm} for more related works and reviews \cite{Altamirano:2014tva,Kubiznak:2016qmn}).

An extraordinary feature of thermodynamics is its universality, as well as gravity. The largest known system controlled by gravity,i.e, the universe, should also obey thermodynamic laws. FRW universe is a dynamical spherically symmetric spacetime and some studies have shown that the existence of the apparent horizon is the main cause of having a self-consistent thermodynamics \cite{Bak:1999hd,Cai:2005ra,Gong:2007md}. Although the FRW universe
has many well-known properties, including the existence of an apparent horizon \cite{Hayward:1993wb,Hayward:1997jp,Jacobson:1995ab}, Hawking temperature and radiation \cite{Cai:2008gw,Hu:2010tx}, Bekenstein-Hawking entropy, quasi-local Misner-Sharp energy \cite{Cai:2009qf,Hu:2015xva}, and the unified first law \cite{Cai:2006rs,Akbar:2006kj,Akbar:2006mq}. However, the thermodynamic equation of state and van der Waals-like phase transition have not yet been known to exist for the FRW universe.

In our recent paper, we investigated the thermodynamics of the FRW universe with a perfect fluid in Einstein gravity and found no existence of a $P$-$V$ phase transition \cite{Abdusattar:2021wfv}. In order to construct a van der Waals-like equation of state that can describe $P$-$V$ phase transitions of the FRW universe, we focused our attention on modified theories of gravity \cite{Kong:2021dqd,Abdusattar:2023hlj} that belongs to the Horndeski class. We obtained a reasonable thermodynamic equation of state for the FRW universe and found a van der Waals-like ($P$-$V$) phase transition. Remarkably, coexistence phase of the $P$-$V$ phase transition occur above the critical temperature, which is different from usual van der Waals system and most of black holes system, providing insights into investigating the microstructure of the FRW universe. As an extension of our work, other related thermodynamic studies such as thermodynamic geometry and the underlying microstructure can be explored for the FRW universe.

To gain insight into the microstructure of a thermodynamic system, it is often essential to investigate the Ruppeiner geometry using thermodynamic fluctuations \cite{Ruppeiner:1995zz,Rup:2010}.
Ruppeiner proposed a line element on the equilibrium thermodynamic state space, which measures the second derivative of entropy. The resulting thermodynamic curvature, known as the Ricci scalar, has been shown to be connected to microscopic interactions, with repulsive (attractive) interactions leading to a positive (negative) curvature \cite{GRup:2012}. This approach has been used to study the microscopic behavior and phase transitions of various black holes \cite{Wei:2015iwa,Zangeneh:2016snh,Dehyadegari:2016nkd,KordZangeneh:2017lgs,Miao:2018fke,Miao:2018qyh,Chen:2018icg,Ghosh:2019pwy,Ghosh:2020kba,Guo:2019oad}. However, in some situations where the entropy and thermodynamic volume of black holes are not independent, the heat capacity at constant volume becomes zero, leading to a divergent Ruppeiner scalar curvature and the information about the associated black hole microstructure to be missing from the thermodynamic geometry.

To avoid this divergence, Ref.\cite{Wei:2019uqg} introduced a normalized thermodynamic curvature that takes into account the heat capacity at constant volume. This approach was applied to investigate the microscopic behavior of charged AdS black holes in an extended phase space, and found that the divergence behavior of thermodynamic curvature is characterized by a dimensionless constant $-1/8$ that is identical to that of van der Waals fluid. Later, this universal property with dimensionless constant $-1/8$ is also confirmed in many of black holes system \cite{Wei:2019yvs,Wei:2019ctz,Yerra:2020oph,Wu:2020fij,Abdusattar:2023xxs,Hu:2020pmr}.

In this study, we utilize the Ruppeiner geometry as a powerful tool to explore the microstructure of the FRW universe via $P$-$V$ phase transition. By taking the temperature and volume as fluctuation variables investigate the normalized thermodynamic curvature of the FRW universe.
We observe that the critical behavior of thermodynamic curvature indicates the existence of a universal critical exponent of $2$ along both the coexistence small and large phases, along with a universal coefficient of $-1/8$, which are analogous to the van der Waals fluid in its critical point regime.
Furthermore, we have also noticed that the coexistence large phase always exhibits a negative thermodynamic curvature, suggesting the presence of attractive interactions among perfect fluids in the FRW universe during the phase transition. However, in the coexistence small phase, there is a combination of repulsive and attractive interactions.

The paper is organised as follows. In Sec. \ref{sec:therm}, we shall make a brief review on thermodynamics and equation of state for the FRW universe in the gravity with a generalized conformal scalar field. To observe the divergence behavior of the thermodynamic curvature, we provide the coexisting volumes of FRW universe during the $P$-$V$ phase transition in Sec. \ref{sec:PV}. In Sec. \ref{Rup}, we shall study the behavior of thermodynamic curvature for the FRW universe with a perfect fluid along coexisting volumes. Finally, we end the paper with conclusions and discussion in Sec. \ref{secfour}.

\section{Review Previous Results: Thermodynamics of the FRW Universe}\label{sec:therm}

In this section, we make a brief review on thermodynamics and equation of state for the FRW universe with a perfect fluid associated with apparent horizon in gravity with a generalized conformal scalar field that belongs to the Horndeski class \cite{Kong:2021dqd}.
We start our study from the introduction of a well defined action nicely argued in \cite{Fernandes:2021dsb}
\begin{eqnarray}\label{eq:actionconfgeneral}
S&=&\int \frac{d^{4} x \sqrt{-g}}{16\pi}\Big[\mathcal{R}-2\Lambda -\beta e^{2\phi}\left(\mathcal{R} + 6(\nabla \phi)^{2}\right)-2\lambda e^{4\phi} \nonumber\\
&&- \alpha \Big(\phi \mathcal{G} - 4 G^{\mu \nu} \nabla_{\mu} \phi \nabla_{\nu} \phi - 4 \square \phi(\nabla \phi)^{2} - 2(\nabla \phi)^{4}\Big)\Big] \nonumber\\
&&+S_{m}\,,
\end{eqnarray}
The action given in Eq.\eqref{eq:actionconfgeneral} encompasses a scalar-tensor theory characterized by constants $\alpha$, $\beta$, and $\lambda$. It includes the Gauss-Bonnet term $\mathcal{G}$, the determinant of the metric tensor $g_{\mu\nu}$ denoted by $g$, $G^{\mu \nu}$ is the Einstein tensor, the Ricci scalar $\mathcal{R}$, and the action $S_m$ associated with the matter field. Here, $\square \equiv \nabla_\mu \nabla^\mu$ represents the covariant d'Alembertian, and $(\nabla \phi)^2 \equiv \nabla_\mu \phi \nabla^\mu \phi$ denotes the square of the covariant derivative of the scalar field $\phi$, where $\nabla_\mu$ is the covariant derivative.

The field equations in this modified gravity are obtained by
\begin{equation}\label{FieldEq}
G_{\mu \nu} + \Lambda g_{\mu \nu} +\alpha \mathcal{H}_{\mu \nu} - \beta e^{2\phi} \mathcal{A}_{\mu \nu}+\lambda e^{4\phi}g_{\mu \nu}=8\pi T_{\mu\nu}^m\,,
\end{equation}
where $T_{\mu\nu}^m$ is the stress-energy tensor of matter, and
\begin{eqnarray}
\mathcal{H}_{\mu\nu} &=& 2G_{\mu \nu} (\nabla\phi)^2+4P_{\mu \alpha \nu \beta}(\nabla^\alpha \phi \nabla^\beta \phi - \nabla^\beta \nabla^\alpha \phi) \nonumber\\
&+&4(\nabla_\alpha \phi \nabla_\mu \phi - \nabla_\alpha \nabla_\mu \phi) (\nabla^\alpha \phi \nabla_\nu \phi - \nabla^\alpha \nabla_\nu \phi)\nonumber\\
&+&4(\nabla_\mu \phi \nabla_\nu \phi - \nabla_\nu \nabla_\mu \phi) \square\phi+g_{\mu \nu} [2(\square\phi)^2 - (\nabla \phi)^4] \nonumber\\
&+&g_{\mu \nu} [2\nabla_\beta \nabla_\alpha \phi (2\nabla^\alpha \phi \nabla^\beta \phi - \nabla^\beta \nabla^\alpha \phi)]\,,\nonumber
\end{eqnarray}
\begin{equation}
\mathcal{A}_{\mu\nu} = G_{\mu \nu} + 2\nabla_\mu \phi \nabla_\nu \phi - 2\nabla_\mu \nabla_\nu \phi +g_{\mu \nu} [2\square\phi + (\nabla\phi)^2] \,,\nonumber
\end{equation}
with
\begin{eqnarray}
P_{\alpha \beta \mu \nu} &\equiv& *R*_{\alpha \beta \mu \nu} = -R_{\alpha \beta \mu \nu}-g_{\alpha \nu} R_{\beta \mu}+g_{\alpha \mu} R_{\beta \nu} \nonumber\\
&&-g_{\beta \mu} R_{\alpha \nu}+g_{\beta \nu} R_{\alpha \mu}-\frac{1}{2}\left(g_{\alpha \mu} g_{\beta \nu}-g_{\alpha \nu} g_{\beta \mu}\right) \mathcal{R}\,.\nonumber
\end{eqnarray}
Remarkably, the particularity of the construction in \cite{Fernandes:2021dsb} lies in the fact that the trace of the metric equations, along with the scalar field equation derived from the action (\ref{eq:actionconfgeneral}), form a purely geometric four-dimensional equation:
\begin{equation}\label{Eq:trace}
\mathcal{R}+\frac{\alpha}{2}\mathcal{G}-4\Lambda = -8\pi T^m \,,
\end{equation}
which bears a striking resemblance to the trace equation found in the higher-dimensional Einstein-Gauss-Bonnet theory, where $T^m=g^{\mu\nu}T_{\mu\nu}^m$ is the trace of the stress-energy tensor.

In the co-moving coordinate system $\{t,r,\theta,\varphi\}$, the line element of the spatially flat FRW universe can be written as
\begin{equation}
d s^2=-d t^2+a^2(t)[d r^2+r^2(d\theta^2+\sin^2\theta d\varphi^2)]\,, \label{LE}
\end{equation}
where $a(t)$ is the time-dependent scale factor. We assume the source of matter field for the FRW universe as the perfect fluid with stress energy momentum tensor
\begin{equation}
T_{\mu\nu}^m=(\rho_m+p_m)u_{\mu}u_{\nu}+p_m g_{\mu\nu}\,,\label{ST}
\end{equation}
where $\rho_m$ and $p_m$ are energy density and pressure of the perfect fluid, $u_{\mu}$ is the four velocity. Applying the field equation (\ref{FieldEq}) to the FRW universe metric (\ref{LE}) with Eqs.(\ref{Eq:trace}) and (\ref{ST}), the modified Friedmann's equations are obtained by\footnote{Notably, the field equation given in Eq.(\ref{FieldEq}) involves three coupling constants ($\alpha, \beta$, and $\lambda$), but the Friedmann's equations simplifies it by considering only one coupling constant ($\alpha$). For more detailed discussion on this, see reference \cite{Fernandes:2021dsb}.
An another additional point to note is that the forms of Friedmann's equations for a flat FRW universe in literatures \cite{Lu:2020iav,Feng:2020duo,Fernandes:2022zrq,Fernandes:2021dsb,Glavan:2019inb} are same, and thus could share similar thermodynamic properties.}
\begin{eqnarray}
\rho&=&\rho_m+\rho_{\Lambda}=\frac{3H^2}{8\pi}(1+\alpha H^2)\,, \label{GBrhom}\\
p&=&p_m+p_{\Lambda}\nonumber\\
&=&-\frac{3H^2}{8\pi}(1+\alpha H^2)-\frac{1+2\alpha H^2}{4\pi}\dot H \label{GBpm}\,,
\end{eqnarray}
where $\rho_{\Lambda}=-p_{\Lambda}=\frac{\Lambda}{8\pi}$.


For later convenience, we rewrite the FRW universe metric (\ref{LE}) with the areal radius $R\equiv a(t)r$ given by
\begin{equation}\label{NewFRW}
d s^2=h_{ab}d x^a d x^b+R^2(d\theta^2+\sin^2\theta d\varphi^2)\,,
\end{equation}
where $a,b=0,1$ with $x^0=t, x^1=r$ and $h_{ab}=[-1,a^2(t)]$. By using the metric (\ref{NewFRW}), it becomes straightforward to determine the apparent horizon of the FRW universe through the solution of $h^{ab}\partial_a R\partial_b R=0$ \cite{Hayward:1993wb}. This for the metric (\ref{NewFRW}) yields the following result \cite{Bak:1999hd,Cai:2005ra}
\begin{equation}
R_A=\frac{1}{H}\,, \label{AH}
\end{equation}
whose time derivative is then
\begin{equation}
\dot{R}_A=-\dot{H}R^2_A \,. \label{dot}
\end{equation}
Use Eqs.(\ref{AH}) and (\ref{dot}), one can express the energy density (\ref{GBrhom}) and the pressure (\ref{GBpm}) in terms of the apparent horizon radius and its time derivative as
\begin{eqnarray}
\rho&=&\frac{3}{8\pi R^2_A}+\frac{3\alpha}{8\pi R^4_A}\,, \label{rhoGB}\\
p&=&-\frac{3}{8\pi R^2_A}-\frac{3\alpha}{8\pi R^4_A}+\Big(1+\frac{2\alpha}{R^2_A}\Big)\frac{\dot{R}_A}{4\pi R^2_A}\,.\,\,\,\,\,\,\label{pmGB}
\end{eqnarray}
Then the work density\footnote{Note that the definition of work density introduced by Hayward, which is an essential quantity for investigating the thermodynamics of dynamical spherically symmetric spacetimes, see more related discussion in literature \cite{Hayward:1993wb,Hayward:1997jp}.} of the FRW universe associated with the perfect fluid obtained by \cite{Kong:2021dqd}
\begin{eqnarray}\label{WD}
W&=&\frac{1}{2}(\rho-p)\nonumber\\
&=&\frac{3}{8\pi R^2_A}+\frac{3\alpha}{8\pi R^4_A}-\frac{\dot{R}_A}{8\pi R^2_A} \Big(1+\frac{2\alpha}{R^2_A}\Big)\,,\,\,\,\,\,
\end{eqnarray}
which will be used in this paper to define the thermodynamic pressure of the FRW universe.


The surface gravity of spatially flat FRW universe at the apparent horizon can be obtained as \cite{Cai:2005ra}
\begin{equation}
\kappa\equiv \left. \frac{1}{2\sqrt{-h}}\partial_a(\sqrt{-h}h^{ab}\partial_b R) \right|_{R=R_{A}}\,,
\end{equation}
which gives
\begin{equation}\label{SurfaceG}
\kappa=-\frac{1}{R_A}\Big(1-\frac{\dot{R}_A}{2}\Big)\,.
\end{equation}
Assuming $\dot{R}_A$ to be a small quantity, the surface gravity $\kappa$ of the apparent horizon in the FRW universe is negative\footnote{See relevant previously works on negative surface gravity in literature \cite{Hayward:1993wb,Abdusattar:2021wfv,Abdusattar:2022bpg,Dolan:2013ft}.}, which should be required associated with the perfect fluid
\begin{equation}\label{nkcondFRWa}
p-\frac{\rho}{3}<\frac{4\rho}{3}\frac{\alpha}{R_A^2+\alpha}\,,
\end{equation}
see a more detailed derivation in the Appendix (\ref{kappaFRWa}).
Then the Hawking temperature associated with the apparent horizon of the spatially flat FRW universe is obtained by
\begin{equation}
T\equiv\frac{|\kappa|}{2\pi}=\frac{1}{2\pi R_A}\Big(1-\frac{\dot{R}_A}{2}\Big)\,. \label{HT}
\end{equation}
It is important to note that, to derive above formula, the assumption $\dot R_A\ll 1$ must be imposed. In fact, throughout this paper our discussion is focused on the scenario of approximate thermal equilibrium, i.e.\ the FRW universe undergoes a very slow evolution, and thus all the thermodynamic processes are quasistatic.

The Misner-Sharp energy of the FRW universe associated with the apparent horizon is given by \cite{Cai:2009qf,Maeda:2007uu,Cai:2008mh,Kong:2021dqd}
\begin{equation}
E\equiv\frac{R_A}{2}+\frac{\alpha}{2R_A}\,.
\end{equation}

By using the above defined quantities, the thermodynamic first law is obtained by
\begin{equation}
d E=-T d S+W d V\,, \label{FL}
\end{equation}
where the entropy\footnote{The first term represents to standard Bekenstein-Hawking entropy, while logarithmic corrections to black hole entropy are often seen as subleading terms in various contexts related to quantum gravity \cite{Sen:2012dw,Sheykhi:2010wm,Zhu:2009qc,Cai:2008ys}.}\cite{Fernandes:2020rpa,Kaul:2000kf,Cai:2009ua,Mukherji:2002de}
\begin{equation}
S= \frac{A}{4}+2\pi \alpha \ln \Big(\frac{A}{A_0}\Big)\,,\label{SS}
\end{equation}
where $A = 4\pi R_A^2$ is the area of the apparent horizon and $A_0$ a constant with units of area,
and also the thermodynamic volume,
\begin{equation}
V\equiv\frac{4}{3}\pi R_A^3\,. \label{V}
\end{equation}
It is essential to note that the apparent simplicity arising from the Eq.(\ref{FL}) is augmented by the conjugate relationships between thermodynamic variables. Specifically, the temperature ($T$) is conjugate to entropy ($S$), and work density ($W$) is conjugate to volume ($V$), wherein both entropy ($S$) and volume ($V$) are explicitly dependent on the apparent horizon radius $R_A$. The rationale for the separation of the terms $T dS$ and $W dV$ in Eq.(\ref{FL}) lies in its ability to distinctly elucidate the energy pathways associated with temperature and work.
For more discussion about the first law of thermodynamics, see the Appendix \ref{B}.


Comparing the Eq.(\ref{FL}) with the standard form of the first law of thermodynamics
\begin{equation}
d U=T d S-P d V\,, \label{SFL}
\end{equation}
one can see that the Misner-Sharp energy $E$ should be interpreted as the minus of the internal energy $U:=-E$, and the work density $W$ as the thermodynamic pressure $P$, i.e.,
\begin{eqnarray}\label{WW}
P:=W\,.
\end{eqnarray}

Using Eqs.(\ref{WD}), (\ref{V}) and (\ref{WW}), one can straightforwardly obtain the thermodynamic equation of state for the FRW universe in modified gravity with a generalized conformal scalar field \cite{Kong:2021dqd}
\begin{equation}\label{EoSPVT}
P=\frac{T}{2}\Big(\frac{4\pi}{3V}\Big)^{\frac{1}{3}}+\frac{1}{8\pi}\Big(\frac{4\pi}{3V}\Big)^{\frac{2}{3}} + \frac{4\pi \alpha T}{3V} -\frac{\alpha}{8\pi}\Big(\frac{4\pi}{3V}\Big)^{\frac{4}{3}}\,.
\end{equation}

\section{Coexistence Curves of the $P$-$V$ Phase Transition}\label{sec:PV}

The necessary condition of the system has $P$-$V$ phase transition is
\begin{equation}
\Big(\frac{\partial P}{\partial V}\Big)_{T}=\Big(\frac{\partial^2 P}{\partial V^2}\Big)_{T}=0\,,\label{PVTc}
\end{equation}
has a critical-point solution $T=T_c,\ P=P_c,\ V=V_c$.
By substituting Eq.(\ref{EoSPVT}) into criticality  condition (\ref{PVTc}), the critical volume, critical temperature and critical pressure for negative $\alpha$ are obtained by
\begin{equation}
V_c=\frac{4\pi R_c^3}{3}\,, ~ T_c=\frac{\sqrt{6+4\sqrt{3}}}{12\pi\sqrt{-\alpha}}\,, ~ P_c=\frac{15+8\sqrt{3}}{-288\pi\alpha}\,,
\end{equation}
where $R_c=\sqrt{-(4\sqrt{3}-6)\alpha}$, and thus there is a van der Waals-like phase transition in the FRW universe \cite{Kong:2021dqd}.

\begin{figure}[h]
\centering
\includegraphics[scale=0.6]{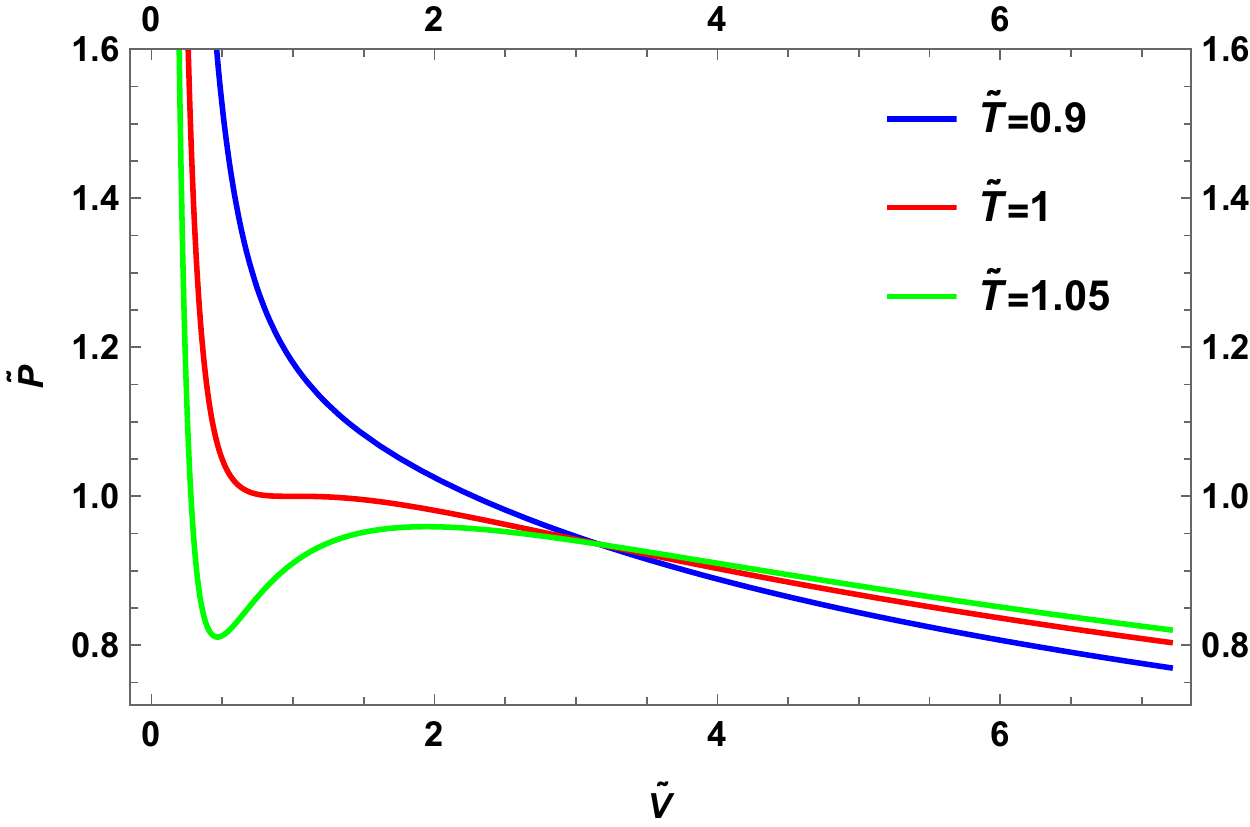}
\caption{$\widetilde{P}$-$\widetilde{V}$ isotherms for the reduced equation of state of FRW universe. It shows an ideal gas behavior indicating a unique phase of FRW universe for $\widetilde{T}<1$, while an oscillatory behavior for $\widetilde{T}>1$, indicating the existence of small and large horizon phases which undergo a first order phase transition that terminates at the critical point $(\widetilde{V}, \widetilde{P}) = (1, 1)$. The isotherms crossover at ($\widetilde{V}_{0}, \widetilde{P}_{\widetilde{V}=\widetilde{V}_{0}}$) with $\widetilde{V}_{0}=\sqrt{45+26\sqrt{3}}/3$, which correspond to thermodynamic singularity.}
\label{FigPV}
\end{figure}

For the purpose of conveniently and explicitly illustrating the phase transition diagram, we define the dimensionless reduced pressure, reduced volume, and reduced temperature as follows
\begin{equation}\label{reduce}
\widetilde{P}\equiv\frac{P}{P_{c}}\,, \quad\quad \widetilde{V}\equiv\frac{V}{V_{c}}\,, \quad\quad \widetilde{T}\equiv\frac{T}{T_{c}}\,,
\end{equation}
and hence rewrite the equation of state (\ref{EoSPVT}) as
\begin{eqnarray}\label{ReducedPVT}
\widetilde{P}&=&\frac{\widetilde{T}}{2}\Big(\frac{4\pi}{3\widetilde{V}}\Big)^{\frac{1}{3}}\frac{T_c}{P_c V_c^{\frac{1}{3}}}+\frac{1}{8\pi}\Big(\frac{4\pi}{3\widetilde{V}}\Big)^{\frac{2}{3}}\frac{1}{P_c V_c^{\frac{2}{3}}}+\frac{4\pi\alpha\widetilde{T}}{3\widetilde{V}}\frac{T_c}{P_c V_c}\nonumber\\
&&-\frac{\alpha}{8\pi}\Big(\frac{4\pi}{3\widetilde{V}}\Big)^{\frac{4}{3}}\frac{1}{P_c V_c^{\frac{4}{3}}}\,,\,\,\,
\end{eqnarray}
and its corresponding critical behavior is shown in Fig.\ref{FigPV}.

By examining Fig.\ref{FigPV}, we can observe that the isotherms for temperatures below the critical temperature $T_c$ display single-phase behavior similar to an ideal gas, indicating a stable state of the FRW universe without any phase transitions. Conversely, the isotherms for temperatures above $T_c$ correspond to the gas-liquid coexistence phase, which undergoes a first-order phase transition. Furthermore, we notice that the phase transition curves gradually shift downwards with increasing reduced temperature $\widetilde{T}$ for $\widetilde{V}<\widetilde{V}_{0}$, and upwards for $\widetilde{V}>\widetilde{V}_{0}$.

To observe the divergence behavior of the thermodynamic curvature in the next section, we provide the reduced  coexisting volumes of FRW universe during the $P$-$V$ phase transition.
For this goal, we make an expansion to equation of state (\ref{EoSPVT}) near the critical point as
\begin{equation}\label{seriesP}
 \widetilde{P}\approx 1+R \tau+B \epsilon \tau +D \epsilon^3+ \mathcal{O}(\tau \epsilon^2, \epsilon^4) \,,
\end{equation}
where $\tau=\widetilde{T}-1$, $\epsilon=\widetilde{V}-1$ and
\begin{eqnarray}
 R&=&-\frac{8}{11}(2 \sqrt{3}-1)\,,~~~~~~~~ B=\frac{8}{33}(5 \sqrt{3}+3)\,,\,\,\nonumber\\ D&=&-\frac{4}{297}(\sqrt{3}+5)\,.\,\,\nonumber
\end{eqnarray}


According to Maxwell's equal area law \cite{Lan:2015bia,Xu:2015hba} (see for more related discussion in Refs.\cite{Spallucci:2013osa,Majhi:2016txt,Bhattacharya:2017hfj}), the oscillatory region is replaced with an isobar, $P^*= P_s=P_l$, cutting the pressure graph in such a way that the area below and above the isobar are the same:
\begin{eqnarray}
\int_s^l Vd{P}=0\,,
\end{eqnarray}
where the labels `s' and `l' represents to `small' and `large' respectively.
Employing Maxwell's equal area law, we obtain the  reduced volumes of the coexistence small and large phases of FRW universe near the critical point, which are
\begin{eqnarray}
\widetilde{V}_s&=&1-\sqrt{2\times{3}^{5/2}(\widetilde{T}-1)}\,,\,\,\,\,\,\label{Vs} \\
\widetilde{V}_l&=&1+\sqrt{2\times
{3}^{5/2}(\widetilde{T}-1)}\,.\,\,\label{Vl}
\end{eqnarray}

\begin{figure}[h]
\centering
\includegraphics[scale=0.6]{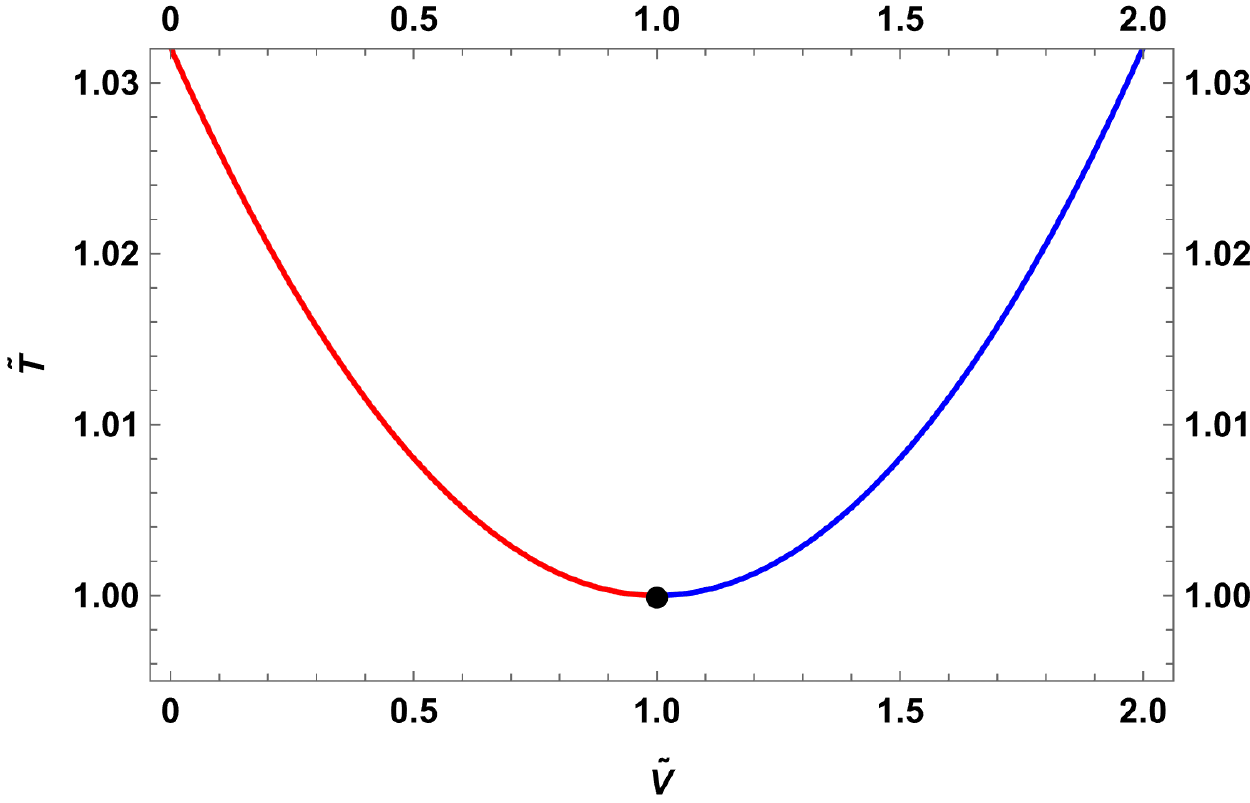}
\caption{Coexistence curves: the reduced volume $\widetilde{V}$ as a function of the reduced temperature $\widetilde{T}$.}
\label{FigTV}
\end{figure}

To gain a better understanding of the phase transition behaviors for the coexistence small and large phases around the critical point, we plot the $\widetilde{V}$-$\widetilde{T}$ diagram corresponding to Eqs.(\ref{Vs}) and (\ref{Vl}), as shown in Fig. \ref{FigTV}.

In Fig. \ref{FigTV}, the red line represents the coexistence curve of small FRW universe phase and the blue line represents the coexistence curve of large FRW universe phase, and they meet at the critical point (black colored point).


\section{Thermodynamic Scalar Curvature of a FRW Universe via $P$-$V$ Criticality}\label{Rup}

The Ruppeiner geometry has proven to be a valuable tool for probing the microscopic structure of thermodynamic systems \cite{Ruppeiner:1995zz}. By examining the thermodynamic curvature, which is connected to microstructure interactions, one can gain a comprehensive understanding of the microstructure of the system\cite{Rup:2010,GRup:2012,Wei:2015iwa}. The sign of thermodynamic curvature determines whether the microstructure interactions are repulsive or attractive: when $R > 0$, repulsive interactions prevail, while $R < 0$ indicates attractive interactions, and $R=0$ signifies no interaction. In this section, we apply Ruppeiner geometry to investigate the thermodynamic scalar curvature of a FRW universe, taking $T$ and $V$ as fluctuation variables.

\subsection{Ruppeiner geometry}\label{RupM}

When studying a thermodynamic system, we typically consider its key properties, such as entropy ($S$), internal energy ($U$), and volume ($V$). The line element between two distinct thermodynamic states can then be elegantly expressed as follows \cite{Ruppeiner:1995zz,Rup:2010}
\begin{equation}
dl^{2}=g_{\mu \nu }d x^{\mu }d x^{\nu }\,,  \label{metric1}
\end{equation}
where $x^{\mu}=\left(U,V\right) $ and the metric element $g_{\mu \nu }$ is given by
\begin{equation}
g_{\mu \nu}=-\frac{\partial ^{2}S}{\partial x^{\mu }\partial x^{\nu}}\,.
\notag
\end{equation}
The first law of thermodynamics for the system in the entropy representation is expressed as follows
\begin{equation}
dS=\frac{1}{T}dU+\frac{P}{T}dV\,,  \label{firstlaw}
\end{equation}
and which leads $({\partial S}/{\partial U})_V={1}/{T}, ({\partial S}/{\partial V})_U={P}/{T}$.
Hence,
\begin{eqnarray}
d\Big(\dfrac{1}{T}\Big)&=&\Big(\dfrac{\partial^2 S}{\partial U^2}\Big)_V dU+\Big(\dfrac{\partial^2 S}{\partial U \partial V}\Big)dV\,,\,\,\\
d\Big(\dfrac{P}{T}\Big)&=&\Big(\dfrac{\partial^2S}{\partial V^2}\Big)_UdV+\Big(\dfrac{\partial^2S}{\partial U\partial V}\Big)dU\,,\,\,
\end{eqnarray}
and the line element $dl^2$ is further equal to
\begin{eqnarray}
dl^2&=&-d\Big(\dfrac{1}{T}\Big)dU-d\Big(\dfrac{P}{T}\Big)dV \nonumber \\
&=&\dfrac{1}{T^2}dTdU-\frac{1}{T}dPdV+\frac{P}{T^2}dTdV\,.\,\,
\end{eqnarray}
Using $dU=C_V dT+\left[T({\partial P}/{\partial T})_V-P\right]dV$ and $dP=({\partial P}/{\partial T})_VdT+({\partial P}/{\partial V})_TdV$ with some simple calculations, one can obtain the thermodynamic line element with $(T,V)$ variables as \cite{Wei:2015iwa,Wei:2019uqg,Wei:2019yvs}
\begin{eqnarray}\label{Newelement}
dl^2=\dfrac{C_V}{T^2}dT^2-\dfrac{(\partial_V P)_T}{T}dV^2\,,
\end{eqnarray}
where the temperature and volume are taken as the fluctuation variables.\footnote{See more related discussions in the literature \cite{Xu:2020gud}.}

\subsection{Thermodynamic curvature via $P$-$V$ criticality}

In this section, our focus is on exploring the thermodynamic geometry of the FRW universe, and revealing insights into its microstructure. However, we encounter a unique characteristic of the FRW universe, where the heat capacity at constant volume ($C_V$) vanishes, i.e. $C_{V}=T\left(\partial S/\partial T\right)_{V}=0$. This is due to the fact that the entropy (\ref{SS}) solely depends on the volume, making the thermodynamic line element (\ref{Newelement}) singular and consequently, it is scalar curvature $R$ will be divergent. As a result, information about the associated microstructure of the FRW universe remains elusive through thermodynamic geometry. To overcome this problem, we introduce a normalized thermodynamic scalar curvature as a means of investigating the thermodynamic properties of the FRW universe. This scalar curvature is defined as follows \cite{Wei:2019uqg,Wei:2019yvs,Yerra:2020oph,Hu:2020pmr}
\begin{eqnarray}\label{NTC}
R_{N}&=&C_{V} R\\
&=& \frac{(\partial_V P)^2 - T^2(\partial_{V, T} P)^2 + 2T^2(\partial_V P)(\partial_{V, T, T} P)}{2(\partial_V P)^2}\,.\nonumber
\end{eqnarray}
In what follows, we analyze the behavior of the normalized thermodynamic scalar curvature for the FRW universe in detail.

By performing simple calculations, the normalized scalar curvature of FRW universe in terms of reduced thermodynamic variables as

\begin{eqnarray}\label{RN}
R_{N}=\frac{2 \Xi \widetilde{T} \widetilde{V}^{5/3}-2 \sqrt{3} \widetilde{T} {\widetilde{V}}^{1/3}-4 \eta \widetilde{T} \widetilde{V}+[1-\beta \widetilde{V}^{2/3}]^2}{2[\sqrt{3} \widetilde{T} {\widetilde{V}}^{1/3}-\eta \widetilde{T} \widetilde{V}+\beta \widetilde{V}^{2/3}-1]^2} \,\,\
\end{eqnarray}
with
\begin{eqnarray}
\Xi=7 \sqrt{3}-12\,,~~~~~
\eta=2-\sqrt{3}\,,~~~~~
\beta=3-2 \sqrt{3}\,.\nonumber
\end{eqnarray}
Note that $R_N$ does not explicitly depend on the gravitational coupling constant $\alpha$ — the system with different $\alpha$ share the same expression in the reduced parameter space, a universal result.
We observe from Eq.(\ref{RN}) that the $R_{N}$ diverges at the temperature
\begin{eqnarray}
\widetilde{T}_{div} =
\frac{1-\beta\widetilde{V}^{2/3}}{\sqrt{3} {\widetilde{V}}^{1/3}-\eta \widetilde{V}}
\,.\,\,\,\label{div1}
\end{eqnarray}

Finally, we discuss the behaviors of thermodynamic curvature towards the critical point along the coexistence curve. For this purpose, we write $R_{N}$ as a function of $\widetilde{T}$ by using Eqs.(\ref{RN}), (\ref{Vs}) and (\ref{Vl}), and the series expansions of it at $\widetilde{T}$ along the coexistence curves have the
following forms
\begin{eqnarray}
R_{N}^{s}&=&-\frac{1}{8 \tau^2}+\frac{\sqrt{-27+42 \sqrt{3}}}{4 \tau^{3/2}}+O(\tau^{-1})\,, \label{RNs}\\
R_{N}^{l}&=&-\frac{1}{8 \tau^2}-\frac{\sqrt{-27+42 \sqrt{3}}}{4 \tau^{3/2}}+O(\tau^{-1})\,, \label{RNl}
\end{eqnarray}
where $\tau=\widetilde{T}-1$ is the deviation from the critical temperature.
We see that $R_{N}\rightarrow - \infty$ at the critical point with a universal critical exponent of $2$.

\begin{figure}[h]
\includegraphics[scale=0.6]{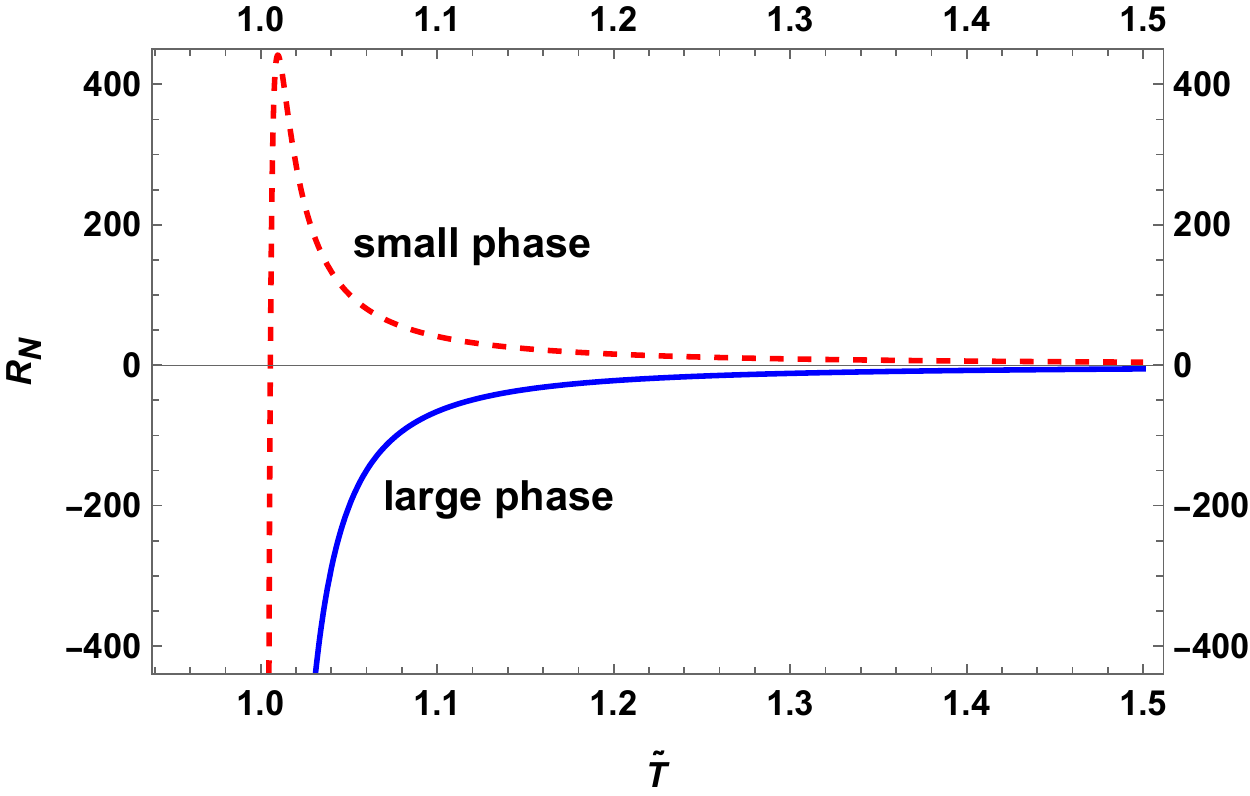}
\caption{The normalized thermodynamic curvature $R_{N}$ of the FRW universe along the coexistence small and large phases. For the small phase, the $R_{N}$ changes the sign to positive at $\widetilde{T}\approx 1.00546$.}
\label{FigRNT}
\end{figure}

Furthermore, by utilizing Eqs.(\ref{RNs}) and (\ref{RNl}) and ignoring the high orders, we obtain the following expression:
\begin{equation}\label{universal C}
 \lim_{\tau\rightarrow 0} R_{N} \tau^2=-\frac{1}{8}\,.
\end{equation}
This reveals a dimensionless universal constant of $-1/8$. Remarkably, this constant obtained analytically, is in exact agreement with the numerical result obtained for the van der Waals fluid \cite{Wei:2019uqg}.
To see the divergence behavior of $R_{N}$, we illustrate the normalized thermodynamic curvature along the coexistence small and large phases in Fig.\ref{FigRNT}.

Figure \ref{FigRNT} shows that the normalized thermodynamic curvature $R_N$ in both small and large phases diverges to negative infinity ($R_N \rightarrow -\infty$) at the critical point. Moreover, $\left\vert R_{N}\right\vert $ decreases as the temperature increases from the critical temperature, becoming small at large $\widetilde{T}$.
Note that the $R_N$ for coexistence large phase  remains negative, indicating attractive microstructure interactions. On the other hand, in the coexistence small phase, $R_N$ changes sign and becomes positive above $\widetilde{T}\approx 1.00546$. In this region, the microstructure interaction of the coexistence small phase in the FRW universe during the phase transition undergo from attractive to repulsive, and the $R_N$ tending to zero as $\widetilde{T}$ grows. In the lower temperature regime (above the critical point), a strongly repulsive interaction along with coexistence small phase is also present.

\section{Conclusions and Discussion}\label{secfour}



In this paper, we have reviewed briefly the $P$-$V$ phase transition of the spatially flat FRW universe in the most general subset of Horndeski theories whose scalar field equation is conformally invariant. In this
study, the thermodynamic pressure identified with the work density of perfect fluid from the first law of thermodynamics to construct an equation of state, while in asymptotical AdS black holes necessarily having the cosmological constant treat as thermodynamic pressure. Impressively, the coexistence phase of $P$-$V$ phase transition in FRW universe occur above the critical temperature, which is significantly different from that of usual van der Waals system and most of AdS black hole system.
We calculated the reduced thermodynamic volumes of the coexistence small and large phases of the FRW universe around the critical point with reduced temperature, and showed the behaviors on the  $\widetilde{V}$-$\widetilde{T}$ plot.



As an extension of our work, we explored the microscopic properties of the FRW universe by investigating the behavior of normalized thermodynamic curvature $R_{N}$ via $P$-$V$ phase transition take the temperature and volume as the fluctuation variables.
We observed that the critical behavior of $R_{N}$ indicates the existence of a universal critical exponent of $2$ along both the coexistence small and large phase curves, along with a universal coefficient of $-1/8$, which are analogous to the van der Waals fluid in its critical point regime.
However, there is a strongly repulsive interaction exists for a coexistence small phase near the critical temperature regime (above the critical temperature) and gradually approaches zero as the temperature increases, which are different from the usual liquid-gas phase transition. Moreover, the $R_N$ for coexistence large phase is always negative which may be related to the information of attractive interaction among the perfect fluid matter constituents of the FRW universe.


We find it intriguing to extend our investigations to other modified theories of gravity that are coupled to other types of matter fields such as scalar fields, whose physical interpretations are still unknown. Additionally, the van der Waals-like ($P$-$V$) phase transition behavior of the FRW universe lacks a suitable holographic interpretation at present, which is an interesting and meaningful topic that deserves further investigation. We also interested in understanding when these phase transitions occur during the evolution of the universe and whether similar phase transitions can be detected through cosmological observations. Our findings establish a theoretical framework for future astronomical observations, and we intend to investigate these open questions in our future research.

%

\appendix
\section{The Negativity Condition of Surface Gravity for the FRW Universe in Gravity with a Generalized Conformal Scalar Field} \label{A}

In this appendix, we prove the equivalence between the expression (\ref{SurfaceG}) for negative surface gravity, i.e., $\kappa|_{R=R_A}<0$, and Eq.(\ref{nkcondFRWa}).

Since
\begin{equation}\label{conditiona}
\frac{\dot{R}_A}{2}<1 \,,
\end{equation}
we can substitute Eqs. (\ref{AH}) and (\ref{dot}) into (\ref{conditiona}) to see that $\kappa|_{R=R_A}<0$ is equivalent to
\begin{equation}
-\frac{R_A^2 \dot{H}}{2}<1 \,,
\end{equation}
which, in turn, using Eqs. (\ref{rhoGB}) and (\ref{pmGB}), leads to
\begin{eqnarray} \label{kappaA1}
\frac{2\pi (\rho+p)R_A^2}{1+2\alpha H^2}<1 \,.
\end{eqnarray}
Thus,
\begin{eqnarray} \label{kappaA2}
\rho+p<\frac{1+2\alpha H^2}{2\pi R_A^2} \,.
\end{eqnarray}
Rearranging (\ref{kappaA2}), we obtain
\begin{eqnarray} \label{kappaA3}
\rho+p<\frac{1}{2\pi R_A^2}\Big(1+\frac{2\alpha}{R_A^2}\Big)\,,
\end{eqnarray}
and using Eq.(\ref{rhoGB}), we can show that (\ref{kappaA3}) is equivalent to
\begin{eqnarray} \label{kappaA4}
\rho+p<\frac{4\rho}{3}\Big(1+\frac{\alpha}{R_A^2 +\alpha}\Big)\,,
\end{eqnarray}
which can be rewritten as
\begin{eqnarray} \label{kappaFRWa}
p-\frac{\rho}{3}<\frac{4\rho}{3}\frac{\alpha}{R_A^2+\alpha} \,,
\end{eqnarray}
as in Eq. (\ref{nkcondFRWa}). In the limit $\alpha\rightarrow0$, the condition equation (\ref{kappaFRWa}) becomes $p-\frac{\rho}{3}<0$, recovering the condition of negative surface gravity for the FRW universe in Einstein gravity \cite{Abdusattar:2021wfv}.
%



\section{The Unified First Law and Properties at the Apparent Horizon of FRW Universe in Gravity with a Generalized Conformal Scalar Field} \label{B}

In this appendix, we initiate our exploration by starting with the unified first law. Our objective is to investigate the behavior of the Friedmann equation at the apparent horizon of the FRW universe in the context of gravity with a generalized conformal scalar field. Our aim is to identify its behavior as a thermodynamic system, adhering to the first law of thermodynamics in the form depicted by the expression (\ref{FL}).

The well-known unified first law \cite{Hayward:1997jp,Cai:2006rs,Hu:2015xva}
\begin{eqnarray} \label{UFdE}
d \widetilde {E}=\widetilde {A} \widetilde {\Psi}_{a}dx^{a}+\widetilde {W} d \widetilde {V}\,,
\end{eqnarray}
where $\widetilde {A}=4\pi R^2$ and $\widetilde {V}={4\pi R^3}/{3}$ are area and volume of the $3$-dimensional sphere with radius $R$. The work density $\widetilde {W}$ and energy-supply $\widetilde {\Psi}_{a}$ are defined as
\begin{equation}
\widetilde {W}:=-\frac{1}{2}h^{a b}T_{a b} ,~~~~~~~ \widetilde {\Psi}_{a}:=T_{a}^b{\partial _{b}R}+\widetilde {W}{\partial _{a}R}\,.
\end{equation}
From the unified first law with Eqs.(\ref{ST}), (\ref{GBrhom}), (\ref{GBpm}), (\ref{NewFRW}), we obtain
\begin{eqnarray}
\widetilde {A}\widetilde {\Psi}&\equiv&\widetilde {A}\widetilde {\Psi}_{a}dx^{a}\nonumber\\
&=&\frac{\widetilde {A}}{2}(\rho+p)(-HRdt+adr) \nonumber\\
&=& -\frac{\widetilde {A}}{2}\frac{1+2\alpha H^2}{4\pi}\dot H(dR-2HRdt)\,.
\end{eqnarray}
At the apparent horizon of the FRW universe, i.e.,$R=R_A$, we obtain
\begin{eqnarray}
A\Psi&=&(\widetilde {A}\widetilde {\Psi})_{R=R_A}
\nonumber \\
&=&\frac{A}{2}\Big(1+\frac{2\alpha}{R^2_A}\Big)\frac{\dot{R}_A}{4\pi R^2_A}(dR_A-2dt) \nonumber\\
&=& \frac{A}{2}\Big(1+\frac{2\alpha}{R^2_A}\Big)\frac{\dot{R}_A}{4\pi R^2_A}dR_A-\frac{A}{2}\Big(1+\frac{2\alpha}{R^2_A}\Big)\frac{d{R}_A}{2\pi R^2_A} \nonumber\\
&=& \frac{\dot{R}_A}{2}\left(1+\frac{2\alpha}{R_A^2}\right)dR_A-\left(1+\frac{2\alpha}{R_A^2}\right)dR_A \nonumber\\
&=& \frac{1}{2\pi R_A}\Big(1-\frac{\dot{R}_A}{2}\Big)\Big(2\pi R_A+\frac{4\pi \alpha}{R_A}\Big)dR_A \nonumber\\
&=& -TdS \,.
\end{eqnarray}
The work density associated with the apparent horizon is formulated as $W = \widetilde {W}_{R=R_A}=\frac{1}{2}(\rho-p)$.
Utilizing the aforementioned derivations, the thermodynamic first law can be universally expressed, as depicted in Eq.(\ref{FL}), where the energy is denoted as $E=\widetilde{E}_{R=R_A}$.
The work term $WdV$ can be interpreted as the work done due to the change of the apparent horizon, while the expression $A\Psi$ signifies the energy flow traversing through the apparent horizon.

\section*{Acknowledgment}

I express my gratitude to the anonymous referee for the careful review of this work and his/her valuable comments, which has led to a significant improvement in the quality of my manuscript.
I would like to thank Prof. Ya-Peng Hu and Dr. Shi-Bei Kong for their helpful discussions. I would also like to express my gratitude to my family for their selfless support and care, love and understanding, which has enabled me to devote my full attention to my research and academic pursuits.



\begin{thebibliography}{10}

\bibitem{Kastor:2009wy}
D.~Kastor, S.~Ray, and J.~Traschen,
Class. Quant. Grav. \textbf{26}, 195011 (2009),
\href{https://arxiv.org/pdf/0904.2765}{[arXiv:hep-th/0904.2765]}.

\bibitem{Dolan:2010ha}
B.~P.~Dolan,
Class. Quant. Grav. \textbf{28}, 125020 (2011),
\href{https://arxiv.org/pdf/1008.5023}{[arXiv:gr-qc/1008.5023]};
B. P. Dolan,
Class. Quant. Grav. \textbf{28}, 235017 (2011),
\href{https://arxiv.org/pdf/1106.6260}{[arXiv:gr-qc/1106.6260]}.

\bibitem{Kubiznak:2012wp}
D.~Kubiznak and R.~B.~Mann,
JHEP \textbf{07}, 033 (2012),
\href{https://arxiv.org/pdf/1205.0559}{[arXiv:hep-th/1205.0559]}.

\bibitem{Hu:2018qsy}
Y.~P.~Hu, H.~A.~Zeng, Z.~M.~Jiang and H.~Zhang,
Phys. Rev. D \textbf{100}, no.8, 084004 (2019),
\href{https://arxiv.org/pdf/1812.09938}{[arXiv:gr-qc/1812.09938]}.


\bibitem{Bhattacharya:2017hfj}
K.~Bhattacharya and B.~R.~Majhi,
Phys.\ Rev.\ D {\bf 95}, no. 10, 104024 (2017),
\href{https://arxiv.org/abs/1702.07174}{[arXiv:gr-qc/1702.07174]}.

\bibitem{Bhattacharya:2017nru}
K.~Bhattacharya, B.~R.~Majhi and S.~Samanta,
Phys. Rev. D \textbf{96}, no.8, 084037 (2017), \href{https://arxiv.org/abs/1709.02650}{[arXiv:gr-qc/1709.02650]}.

\bibitem{Hendi:2012um}
S.~H.~Hendi and M.~H.~Vahidinia,
Phys. Rev. D \textbf{88}, no.8, 084045 (2013),
\href{https://arxiv.org/abs/1212.6128}{[arXiv:hep-th/1212.6128]}.

\bibitem{Hendi:2017fxp}
S.~H.~Hendi, R.~B.~Mann, S.~Panahiyan and B.~Eslam Panah,
Phys. Rev. D \textbf{95}, no.2, 021501 (2017),
\href{https://arxiv.org/pdf/1812.09938}{[arXiv:gr-qc/1702.00432]}.

\bibitem{Gunasekaran:2012dq}
S.~Gunasekaran, R.~B.~Mann and D.~Kubiznak,
JHEP \textbf{11}, 110 (2012),
\href{https://arxiv.org/abs/1208.6251}{[arXiv:hep-th/1208.6251]}.

\bibitem{Wei:2012ui}
S.~W.~Wei and Y.~X.~Liu,
Phys. Rev. D \textbf{87}, no.4, 044014 (2013),
\href{https://arxiv.org/abs/1209.1707}{[arXiv:gr-qc/1209.1707]}.

\bibitem{Cai:2013qga}
R.~G.~Cai, L.~M.~Cao, L.~Li and R.~Q.~Yang,
JHEP \textbf{09}, 005 (2013),
\href{https://arxiv.org/abs/1306.6233}{[arXiv:gr-qc/1306.6233]}.

\bibitem{Dehghani:2014caa}
M.~H.~Dehghani, S.~Kamrani and A.~Sheykhi,
Phys. Rev. D \textbf{90} (2014) no.10, 104020,
\href{https://arxiv.org/abs/1505.02386}{[arXiv:hep-th/1505.02386]}.

\bibitem{Xu:2015rfa}
J.~Xu, L.~M.~Cao and Y.~P.~Hu,
Phys. Rev. D \textbf{91}, no.12, 124033 (2015),
\href{https://arxiv.org/pdf/1506.03578}{[arXiv:gr-qc/1506.03578]}.

\bibitem{Cai:2014znn}
R.~G.~Cai, Y.~P.~Hu, Q.~Y.~Pan and Y.~L.~Zhang,
Phys. Rev. D \textbf{91}, no.2, 024032 (2015), \href{https://arxiv.org/abs/1409.2369}{[arXiv:hep-th/1409.2369]}.

\bibitem{Spallucci:2013osa}
E.~Spallucci and A.~Smailagic,
Phys. Lett. B \textbf{723}, 436-441 (2013),
\href{https://arxiv.org/abs/1305.3379}{[arXiv:hep-th/1305.3379]}.

\bibitem{Majhi:2016txt}
B.~R.~Majhi and S.~Samanta,
Phys.\ Lett.\ B {\bf 773}, 203 (2017),
\href{https://arxiv.org/abs/1609.06224}{[arXiv:gr-qc/1609.06224]}.

\bibitem{Cheng:2016bpx}
P.~Cheng, S.~W.~Wei and Y.~X.~Liu,
Phys. Rev. D \textbf{94} (2016), 024025,
\href{https://arxiv.org/abs/1603.08694}{[arXiv:gr-qc/1603.08694]}.

\bibitem{Dehyadegari:2017hvd}
A.~Dehyadegari and A.~Sheykhi,
Phys. Rev. D \textbf{98}, no.2, 024011 (2018),
\href{https://arxiv.org/abs/1711.01151}{[arXiv:gr-qc/1711.01151]}.

\bibitem{Kumara:2019xgt}
A.~N. Kumara, C.~L.~A. Rizwan, D.~Vaid and K.~M. Ajith,
  \href{https://arxiv.org/abs/1906.11550}{[arXiv:gr-qc/1906.11550]}.


\bibitem{Li:2020xkh}
R.~Li and J.~Wang,
Phys. Lett. B \textbf{813}, 136035 (2021),
\href{https://arxiv.org/abs/2009.09319}{[arXiv:gr-qc/2009.09319]}.

\bibitem{Abdusattar:2023fdm}
H.~Abdusattar,
\href{https://inspirehep.net/literature/2677214}{Eur. Phys. J. C \textbf{83}, no.7, 614 (2023)}.


\bibitem{Altamirano:2014tva}
N.~Altamirano, D.~Kubiznak, R.~B.~Mann and Z.~Sherkatghanad,
Galaxies \textbf{2} (2014), 89-159,
\href{https://arxiv.org/abs/1401.2586}{[arXiv:hep-th/1401.2586]}.

\bibitem{Kubiznak:2016qmn}
D.~Kubiznak, R.~B.~Mann and M.~Teo,
Class. Quant. Grav. \textbf{34}, no.6, 063001 (2017),
\href{https://arxiv.org/abs/1608.06147}{[arXiv:hep-th/1608.06147]}.


\bibitem{Bak:1999hd}
D.~Bak and S.~J.~Rey,
Class. Quant. Grav. \textbf{17}, L83 (2000),
\href{https://arxiv.org/abs/hep-th/9902173}{[arXiv:hep-th/9902173]}.

\bibitem{Cai:2005ra}
R.~G.~Cai and S.~P.~Kim,
JHEP \textbf{02}, 050 (2005),
\href{https://arxiv.org/abs/hep-th/0501055v1}{[arXiv:hep-th/0501055]}.

\bibitem{Gong:2007md}
Y.~Gong and A.~Wang,
Phys. Rev. Lett. \textbf{99} (2007), 211301,
\href{https://arxiv.org/abs/0704.0793}{[arXiv:hep-th/0704.0793]}.

\bibitem{Hayward:1993wb}
S.~A.~Hayward,
\href{https://inspirehep.net/literature/358106}{Phys. Rev. D \textbf{49}, 6467-6474 (1994)}.

\bibitem{Hayward:1997jp}
S.~A.~Hayward,
Class. Quant. Grav. \textbf{15}, 3147-3162 (1998),
\href{https://arxiv.org/abs/gr-qc/9710089}{[arXiv:gr-qc/9710089]}.

\bibitem{Jacobson:1995ab}
T.~Jacobson,
Phys. Rev. Lett. \textbf{75}, 1260-1263 (1995),
\href{https://arxiv.org/pdf/gr-qc/9504004}{[arXiv:gr-qc/9504004]}.

\bibitem{Cai:2008gw}
R.~G.~Cai, L.~M.~Cao and Y.~P.~Hu,
Class. Quant. Grav. \textbf{26} (2009), 155018,
\href{https://arxiv.org/abs/0809.1554}{[arXiv:hep-th/0809.1554]}.

\bibitem{Hu:2010tx}
Y.~P.~Hu,
Phys. Lett. B \textbf{701} (2011), 269-274,
\href{https://arxiv.org/abs/1007.4044}{[arXiv:gr-qc/1007.4044]}.


\bibitem{Cai:2009qf}
R.~G.~Cai, L.~M.~Cao, Y.~P.~Hu, and N.~Ohta,
Phys. Rev. D \textbf{80} (2009), 104016,
\href{https://arxiv.org/abs/0910.2387}{[arXiv:hep-th/0910.2387]}.

\bibitem{Hu:2015xva}
Y.~P.~Hu and H.~Zhang,
Phys. Rev. D {\bf 92}, no.2, 024006 (2015),
\href{https://arxiv.org/abs/1502.00069}{[arXiv:hep-th/1502.00069]}.


\bibitem{Cai:2006rs}
R.~G.~Cai and L.~M.~Cao,
Phys. Rev. D \textbf{75}, 064008 (2007),
\href{https://arxiv.org/pdf/gr-qc/0611071}{[arXiv:gr-qc/0611071]}.


\bibitem{Akbar:2006kj}
M.~Akbar and R.~G.~Cai,
Phys. Rev. D \textbf{75}, 084003 (2007),
\href{https://arxiv.org/pdf/hep-th/0609128}{[arXiv:hep-th/0609128]}.

\bibitem{Akbar:2006mq}
M.~Akbar and R.~G.~Cai,
Phys. Lett. B \textbf{648}, 243-248 (2007),
\href{https://arxiv.org/abs/gr-qc/0612089}{[arXiv:gr-qc/0612089]}.


\bibitem{Abdusattar:2021wfv}
H.~Abdusattar, S.~B.~Kong, W.~L.~You, H.~Zhang and Y.~P.~Hu,
JHEP \textbf{12}, 168 (2022),
\href{https://arxiv.org/abs/2108.09407}{[arXiv:gr-qc/2108.09407]}.


\bibitem{Kong:2021dqd}
S.~B.~Kong, H.~Abdusattar, Y.~Yin, H.~Zhang and Y.~P.~Hu,
Eur. Phys. J. C \textbf{82}, no.11, 1047 (2022),
\href{https://arxiv.org/abs/2108.09411}{[arXiv:gr-qc/2108.09411]}.


\bibitem{Abdusattar:2023hlj}
H.~Abdusattar, S.~B.~Kong, H.~Zhang and Y.~P.~Hu,
\href{https://arxiv.org/abs/2301.01938}{[arXiv:gr-qc/2301.01938]}.


\bibitem{Ruppeiner:1995zz}
G.~Ruppeiner,
Rev. Mod. Phys. \textbf{67}, 605-659 (1995),
\href{https://journals.aps.org/rmp/abstract/10.1103/RevModPhys.67.605}{[erratum: Rev. Mod. Phys. \textbf{68}, 313-313 (1996)]};
G. Ruppeiner,
\href{https://journals.aps.org/rmp/abstract/10.1103/RevModPhys.67.605}{Phys. Rev. A \textbf{ 20} (1979) 1608}.

\bibitem{Rup:2010}
G. Ruppeiner,
American Journal of Physics, \textbf{78} (2010) 1170,
\href{https://arxiv.org/abs/1007.2160v3}{[arXiv:1007.2160]};
G. Ruppeiner,
Phys. Rev. E\textbf{ 86} (2012) 021130,\href{https://arxiv.org/abs/1208.3265}{[arXiv:1208.3265]}.

\bibitem{GRup:2012}
G. Ruppeiner,
J. Phys.: Conf. Series \textbf{410} (2013) 012138,
\href{https://arxiv.org/abs/1210.2011}{[arXiv:1210.2011]};
G. Ruppeiner,
Springer Proc. Phys. \textbf{153} (2014) 179,
\href{https://arxiv.org/abs/1309.0901}{[arXiv:1309.0901]}.

\bibitem{Wei:2015iwa}
S.~W.~Wei and Y.~X.~Liu,
Phys. Rev. Lett. \textbf{115}, no.11, 111302 (2015)
[erratum: Phys. Rev. Lett. \textbf{116}, no.16, 169903 (2016)],
\href{https://arxiv.org/abs/1502.00386}{[arXiv:gr-qc/1502.00386]}.

\bibitem{Zangeneh:2016snh}
M.~K.~Zangeneh, A.~Dehyadegari and A.~Sheykhi,
\href{https://arxiv.org/abs/1602.03711}{[arXiv:hep-th/1602.03711]}.

\bibitem{Dehyadegari:2016nkd}
A.~Dehyadegari, A.~Sheykhi and A.~Montakhab,
  Phys. Lett. B {\bfseries 768} (2017) 235-240,
\href{https://arxiv.org/abs/1607.05333}{[arXiv:gr-qc/1607.05333]}.

\bibitem{KordZangeneh:2017lgs}
M.~Kord~Zangeneh, A.~Dehyadegari, A.~Sheykhi and R.~B. Mann,
 Phys. Rev. D {\bfseries 97} (2018) 084054,
 \href{https://arxiv.org/abs/1709.04432}{[arXiv:gr-qc/1709.04432]}.

\bibitem{Miao:2018fke}
Y.~G.~Miao and Z.~M.~Xu,
Phys. Rev. D \textbf{98}, no.8, 084051 (2018),
\href{https://arxiv.org/abs/1806.10393}{[arXiv:hep-th/1806.10393]}.


\bibitem{Miao:2018qyh}
Y.-G. Miao and Z.-M. Xu,
Sci. China Phys. Mech. Astron. {\bfseries 62} (2019) 10412,
\href{https://arxiv.org/abs/1804.01743}{[arXiv:hep-th/1804.01743]}.

\bibitem{Chen:2018icg}
Y.~Chen, H.~Li and S.-J. Zhang,
Nucl. Phys. B {\bfseries 948} (2019) 114752,
\href{https://arxiv.org/abs/1812.11765}{[arXiv:hep-th/1812.11765]}.

\bibitem{Ghosh:2019pwy}
A.~Ghosh and C.~Bhamidipati,
  Phys. Rev. D {\bfseries 101} (2020) 046005,
  \href{https://arxiv.org/abs/1911.06280}{[arXiv:gr-qc/1911.06280]}.

\bibitem{Ghosh:2020kba}
A.~Ghosh and C.~Bhamidipati,
Phys. Rev. D \textbf{101} (2020) no.10, 106007,
\href{https://arxiv.org/abs/2001.10510}{[arXiv:hep-th/2001.10510]}.

\bibitem{Guo:2019oad}
X.-Y. Guo, H.-F. Li, L.-C. Zhang and R.~Zhao,
  Phys. Rev. D {\bfseries 100} (2019) 064036,
\href{https://arxiv.org/abs/1901.04703}{[arXiv:gr-qc/1901.04703]}.


\bibitem{Wei:2019uqg}
S.~W.~Wei, Y.~X.~Liu and R.~B.~Mann,
Phys. Rev. Lett. \textbf{123}, no.7, 071103 (2019),
\href{https://arxiv.org/abs/1906.10840}{[arXiv:gr-qc/1906.10840]}.

\bibitem{Wei:2019yvs}
S.~W.~Wei, Y.~X.~Liu and R.~B.~Mann,
Phys. Rev. D \textbf{100}, no.12, 124033 (2019),
\href{https://arxiv.org/abs/1909.03887}{[arXiv:gr-qc/1909.03887]}.

\bibitem{Wei:2019ctz}
S.~W.~Wei and Y.~X.~Liu,
Phys. Lett. B \textbf{803}, 135287 (2020),
\href{https://arxiv.org/abs/1910.04528}{[arXiv:gr-qc/1910.04528]}.

\bibitem{Yerra:2020oph}
P.~K.~Yerra and C.~Bhamidipati,
Int. J. Mod. Phys. A \textbf{35}, no.22, 2050120 (2020),
\href{https://arxiv.org/abs/2006.07775}{[arXiv:hep-th/2006.07775]}.

\bibitem{Wu:2020fij}
B.~Wu, C.~Wang, Z.~M.~Xu and W.~L.~Yang,
Eur. Phys. J. C \textbf{81}, no.7, 626 (2021),
\href{https://arxiv.org/abs/2006.09021}{[arXiv:gr-qc/2006.09021]}.

\bibitem{Abdusattar:2023xxs}
H.~Abdusattar,
\href{https://www.sciencedirect.com/science/article/pii/S2212686423000626?via}{Phys. Dark Univ. \textbf{40}, 101228 (2023)}.

\bibitem{Hu:2020pmr}
Y.~P.~Hu, L.~Cai, X.~Liang, S.~B.~Kong and H.~Zhang,
Phys. Lett. B \textbf{822}, 136661 (2021), \href{https://arxiv.org/abs/2010.09363}{[arXiv:gr-qc/2010.09363]}.




\bibitem{Fernandes:2021dsb}
P.~G.~S.~Fernandes,
Phys. Rev. D \textbf{103} (2021) no.10, 104065,
\href{https://journals.aps.org/prd/abstract/10.1103/PhysRevD.103.104065}{[arXiv:gr-qc/2105.04687]}.


\bibitem{Lu:2020iav}
H.~Lu and Y.~Pang,
Phys. Lett. B \textbf{809} (2020), 135717,
\href{https://arxiv.org/abs/2003.11552}{[arXiv:gr-qc/2003.11552]}.


\bibitem{Feng:2020duo}
J.~X.~Feng, B.~M.~Gu and F.~W.~Shu,
Phys. Rev. D \textbf{103}, 064002 (2021),
\href{https://arxiv.org/abs/2006.16751}{[arXiv:gr-qc/2006.16751]}.


\bibitem{Fernandes:2022zrq}
P.~G.~S.~Fernandes, P.~Carrilho, T.~Clifton and D.~J.~Mulryne,
Class. Quant. Grav. \textbf{39} (2022) no.6, 063001,
\href{https://arxiv.org/abs/2202.13908}{[arXiv:gr-qc/2202.13908]}.


\bibitem{Glavan:2019inb}
D.~Glavan and C.~Lin,
Phys. Rev. Lett. \textbf{124}, no.8, 081301 (2020),
\href{https://arxiv.org/abs/1905.03601}{[arXiv:gr-qc/1905.03601]}.



\bibitem{Abdusattar:2022bpg}
H.~Abdusattar, S.~B.~Kong, Y.~Yin and Y.~P.~Hu,
JCAP \textbf{08}, no.08, 060 (2022),
\href{https://arxiv.org/abs/2203.10868}{[arXiv:gr-qc/2203.10868]}.

\bibitem{Dolan:2013ft}
B.~P.~Dolan, D.~Kastor, D.~Kubiznak, R.~B.~Mann, and J.~Traschen,
Phys. Rev. D \textbf{87}, no.10, 104017 (2013),
\href{https://arxiv.org/pdf/1301.5926.pdf}{[arXiv:hep-th/1301.5926]}.



\bibitem{Maeda:2007uu}
H.~Maeda and M.~Nozawa,
Phys. Rev. D \textbf{77} (2008), 064031,
\href{https://arxiv.org/abs/0709.1199}{[arXiv:hep-th/0709.1199]}.

\bibitem{Cai:2008mh}
R.~G.~Cai, L.~M.~Cao, Y.~P.~Hu and S.~P.~Kim,
Phys. Rev. D \textbf{78} (2008), 124012,
\href{https://arxiv.org/abs/0810.2610}{[arXiv:hep-th/0810.2610]}



\bibitem{Fernandes:2020rpa}
P.~G.~S.~Fernandes,
Phys. Lett. B \textbf{805} (2020), 135468,
\href{https://arxiv.org/abs/2003.05491}{[arXiv:gr-qc/2003.05491]}.


\bibitem{Kaul:2000kf}
R.~K.~Kaul and P.~Majumdar,
Phys. Rev. Lett. \textbf{84}, 5255-5257 (2000),
\href{https://arxiv.org/abs/gr-qc/0002040}{[arXiv:gr-qc/0002040]}.

\bibitem{Cai:2009ua}
R.~G.~Cai, L.~M.~Cao and N.~Ohta,
JHEP \textbf{04}, 082 (2010),
\href{https://arxiv.org/abs/0911.4379}{[arXiv:hep-th/0911.4379]}.

\bibitem{Mukherji:2002de}
S.~Mukherji and S.~S.~Pal,
JHEP \textbf{05}, 026 (2002),
\href{https://arxiv.org/abs/hep-th/0205164}{[arXiv:hep-th/0205164]}.

%
%
%
%
%

\bibitem{Sen:2012dw}
A.~Sen,
JHEP \textbf{04} (2013), 156,
\href{https://arxiv.org/abs/1205.0971}{[arXiv:hep-th/1205.0971]}.


\bibitem{Sheykhi:2010wm}
A.~Sheykhi,
Phys. Rev. D \textbf{81} (2010), 104011,
\href{https://arxiv.org/abs/1004.0627}{[arXiv:gr-qc/1004.0627]};
A.~Sheykhi,
Eur. Phys. J. C \textbf{69} (2010), 265-269,
\href{https://arxiv.org/abs/1012.0383}{[arXiv:hep-th/1012.0383]}.

\bibitem{Zhu:2009qc}
T.~Zhu, J.~R.~Ren and M.~F.~Li,
JCAP \textbf{08} (2009), 010,
\href{https://arxiv.org/abs/0905.1838}{[arXiv:hep-th/0905.1838]}.

\bibitem{Cai:2008ys}
R.~G.~Cai, L.~M.~Cao and Y.~P.~Hu,
JHEP \textbf{08}, 090 (2008),
\href{https://arxiv.org/abs/0807.1232}{[arXiv:hep-th/0807.1232]}.


\bibitem{Xu:2020gud}
Z.~M.~Xu, B.~Wu and W.~L.~Yang,
Phys. Rev. D \textbf{101}, no.2, 024018 (2020),
\href{https://arxiv.org/abs/1910.12182}{[arXiv:gr-qc/1910.12182]}.

\bibitem{Lan:2015bia}
S.~Q.~Lan, J.~X.~Mo and W.~B.~Liu,
Eur. Phys. J. C \textbf{75}, no.9, 419 (2015),
\href{https://arxiv.org/abs/1503.07658}{[arXiv:gr-qc/1503.07658]}.

\bibitem{Xu:2015hba}
H.~Xu and Z.~M.~Xu,
Int. J. Mod. Phys. D \textbf{26}, no.04, 1750037 (2016),
\href{https://arxiv.org/abs/1510.06557}{[arXiv:gr-qc/1510.06557]}.


%


\end{thebibliography}

\end{document}